\documentclass[a4paper]{spie}  %>>> use this instead for A4 paper
\usepackage[]{graphicx}
\usepackage{amssymb,amsmath,psfrag,url}
\newcommand{\rnum}{{\bf {R}}}

\newcommand{\Alt}{{\mathrm{Alt}}}

\newcommand{\VField}[1]{{\bf #1}} 
\newcommand{\MyField}[1]{{\bf #1}}

\def\squareforqed{\hbox{\rlap{$\sqcap$}$\sqcup$}}
\def\qed{\ifmmode\else\unskip\quad\fi\squareforqed}

\title{Finite Element Simulation of Light Propagation in Non-Periodic Mask Patterns}
\author{
Lin Zschiedrich\supit{\,a,b},
Frank Schmidt\supit{\,a,b}
\skiplinehalf
\supit{a}
Zuse Institute Berlin,
Takustra{\ss}e 7,
D\,--\,14\,195 Berlin,
Germany
\smallskip\\
\supit{b}
JCMwave GmbH,
Haarer Stra{\ss}e 14a,
D\,--\,85\,640 Putzbrunn, 
Germany
}

\authorinfo{
Corresponding author: L. Zschiedrich\\
URL: http://www.zib.de/Numerik/NanoOptics/\\
Email: zschiedrich@zib.de
}

\begin{document} 
\maketitle 
%\today
%%%%%%%%%%%%%%%%%%%%%%%%%%%%%%%%%%%%%%%%%%%%%%%%%%%%%%%%%%%%% 
%% SPIE Copyright form 
\noindent
Copyright 2008  Society of Photo-Optical Instrumentation Engineers.\\
This paper will be published in Proc.~SPIE Vol. {\bf 7028}
(2008),  
({\it Photomask and Next Generation Lithography Mask Technology, Toshiyuki Horiuchi, Ed.})
and is made available 
as an electronic preprint with permission of SPIE. 
One print or electronic copy may be made for personal use only. 
Systematic or multiple reproduction, distribution to multiple 
locations via electronic or other means, duplication of any 
material in this paper for a fee or for commercial purposes, 
or modification of the content of the paper are prohibited.
%%%%%%%%%%%%%%%%%%%%%%%%%%%%%%%%%%%%%%%%%%%%%%%%%%%%%%%%%%%%% 
%%%%%%%%%%%%%%%%%%%%%%%%%%%%%%%%%%%%%%%%%%%%%%%%%%%%%%%%%%%%% 
\begin{abstract}
Rigorous electromagnetic field simulations are an essential part for scatterometry and mask pattern design. Today mainly periodic structures are considered in simulations. Non-periodic structures are typically modeled by large, artificially periodified computational domains. For systems with a large radius of influence this leads to very large computational domains to keep the error sufficiently small. In this paper we review recent advances in the rigorous simulation of isolated structures embedded into a surrounding media. We especially address the situation of a layered surrounding media (mask or wafer) with additional infinite inhomogeneities such as resist lines. Further we detail how to extract the far field information needed for the aerial image computation in the non-periodic setting. 
\end{abstract}
\section{Introduction}
\label{sec:Introduction}
\begin{figure}
 \psfrag{Einc}{$\MyField{E}_{\mathrm{inc}}$}
 \psfrag{Esc}{$\MyField{E}_{\mathrm{sc}}$}
 \psfrag{x}{$x$}
 \psfrag{y}{$y$}
 \psfrag{z}{$z$}
  \begin{center}
      \includegraphics[width=6cm]{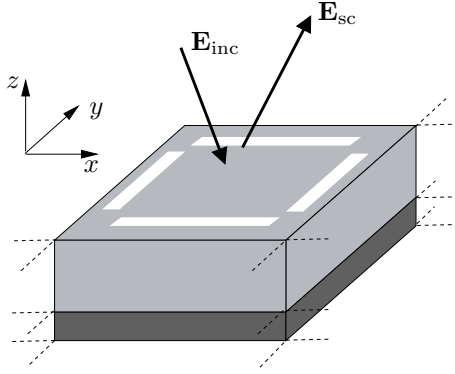}
  \end{center}
  \caption{\label{Fig:TypicalStructure} Sketch of the simulation setting. An incident field $\MyField{E}_{\mathrm{inc}}$ enters the computational domain from above and excites a scattered field $\MyField{E}_{\mathrm{sc}}.$ The computational domain encloses the scatterer. The surrounding media is typically layered. More general situations are depicted in Figure~\ref{Fig:MoreComplicatedGeometries}. 
}
\end{figure}

The simulation setting we focus on is as depicted in Figure~\ref{Fig:TypicalStructure}. An incident field $\MyField{E}_{\mathrm{inc}}$ enters the computational domain from above and excites a scattered field $\MyField{E}_{\mathrm{sc}}.$ For simplicity we assume that the incident field is a plane wave. A more complex illumination can be modeled by a decomposition of the source field into plane waves. Afterwards the scattered fields of the various plane waves are superimposed accordingly to their mutual coherence relation. In Figure~\ref{Fig:TypicalStructure} the scatterer is embedded into a layered media and completely contained in the computational domain. Using a non-periodic domain allows for the reduction of the computational domain to the minimum dimensions of the considered structure. Furthermore, no cutoff error is introduced in contrast to  modeling the structure as a large, artificially periodified computational domain. However, the numerics is more delicate for non-periodic computational domains, since special transparent boundary conditions must be implemented. Besides the more involved numerics for the near field computation the extraction method of the far field data must also account for the non-periodic setting. In a non-periodic direction the diffraction modes are not discrete anymore. Instead we must deal with a band-limited continuous Fourier transform. 

\begin{figure}
 \psfrag{Einc}{$\MyField{E}_{\mathrm{inc}}$}
 \psfrag{Esc}{$\MyField{E}_{\mathrm{sc}}$}
 \psfrag{x}{$x$}
 \psfrag{y}{$y$}
 \psfrag{z}{$z$}
  \begin{center}
      \includegraphics[width=12cm]{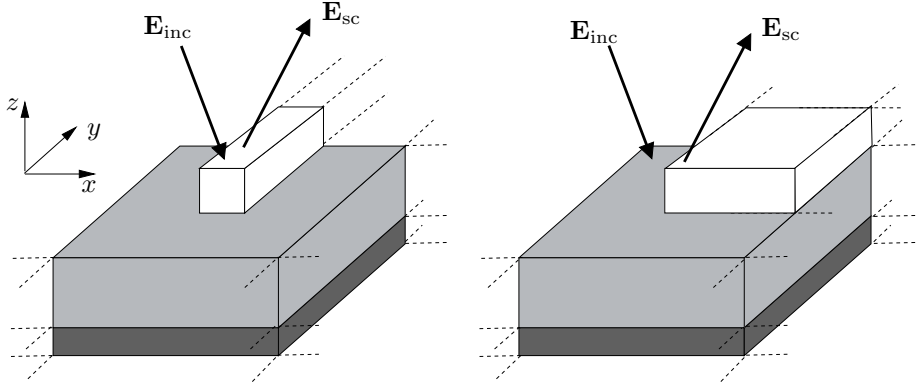}
  \end{center}
  \caption{\label{Fig:MoreComplicatedGeometries} More complicated geometries. We aim to compute the field scattered off an infinite resist line's tip (left hand side) or off an infinite plate's corner. Periodicity in one coordinate direction may also be incorporated but is not especially treated in this paper. 
}
\end{figure}

In earlier papers \cite{Zschiedrich2006a, Burger:2007a} we already studied the
situation as in Figure~\ref{Fig:TypicalStructure}. In this paper we address
the more complicated situation as in
Figure~\ref{Fig:MoreComplicatedGeometries}. Now, the scatterer is not
completely enclosed by the computational domain but is itself infinitely
prolonged as indicated by the dashed lines. Such configurations are called
``waveguide inhomogeneities''~\cite{Schmidt:02a,Zschiedrich:06b}. The theory
presented in our papers~\cite{Schmidt:02a,Zschiedrich:06b} assumes that the
incoming field is zero on the infinite waveguides outside the computational
domain. Therefore, dealing with incident plane waves necessitates a
non-trivial extension of the scattering problem formulation which is presented
in the next section. The presented theory generalizes classical results by
Wiener and Hopf for ideal geometries like a thin half-screen, see~\cite{Titchmarsh:37a,Born:93a}.Accordingly, a new far field extraction formula must be derived for this setting. Numerical aspects on this novel approach are discussed in Section~\ref{Sec:FEMPML}.   
\section{Electromagnetic Scattering Problems} 
\label{Sec:ScatteringProblems}
\subsection{Maxwell's equations}
Starting from Maxwell's equations in a medium without sources and free currents and assuming time-harmonic dependence with angular frequency $\omega>0$ the electric and magnetic fields
\[
\VField{E}(x, y, z, t)  = \widetilde{\VField{E}}(x, y, z)e^{-i\omega \cdot t}, \;
\VField{H}(x, y, z, t) = \widetilde{\VField{H}}(x, y, z)e^{-i\omega \cdot t}, \;
\]
must satisfy 
\begin{eqnarray*}
\nabla \times \widetilde{\VField{E}} & = & i\omega \mu \widetilde{\VField{H}}, \quad
\nabla \cdot \epsilon \widetilde{\VField{E}} = 0, \\
\nabla \times \widetilde{\VField{H}} & = & -i\omega \epsilon \widetilde{\VField{E}}, \quad 
\nabla \cdot \mu \widetilde{\VField{H}} = 0. \\
\end{eqnarray*}
Here $\epsilon$ denotes the permittivity tensor and $\mu$ denotes the permeability tensor of the materials. In the following we drop the wiggles, so that $\widetilde{\VField{E}} \rightarrow \VField{E}$, $\widetilde{\VField{H}} \rightarrow \VField{H}$. From the equations above we then may derive (by direct substitution) the second order equations for the electric field
\begin{eqnarray*}
\nabla \times \mu^{-1} \nabla \times \VField{E} - \omega^2 \epsilon \VField{E} & = & 0, \\
\nabla \cdot \epsilon \VField{E} & = & 0.
\end{eqnarray*}  
Similar equations hold true for the magnetic field - one only need to replace $\VField{E}$ by $\VField{H}$ and interchange $\epsilon$ and $\mu$. Observe that any solution to the first equation also meets the divergence condition (second equation). We therefore drop the second equation in the following.\\
For the sake of a simpler and more compact notation we rewrite these equations in differential form,
\begin{eqnarray}
\label{thefieldmaxequationsdf}
d_{1} \mu^{-1}  d_{1} \VField{e} - \omega^2 \epsilon \VField{e} & = & 0.
\end{eqnarray}  
A reader not familiar with this calculus may replace the exterior derivatives $d_{0}$, $d_{1}$, $d_{2}$ with classical differential operators, $d_{0} \rightarrow \nabla$, $d_{1} \rightarrow \nabla \times $ and $d_{2} \rightarrow \nabla \cdot$. Here, the electric field appears as a differential 1-form, $e = e_{x}dx+e_{y}dy+e_{z}dz,$ whereas the material tensors act -- from a more mathematical point of view -- as operators from the differential 1-forms to the differential 2-forms,
\begin{eqnarray*}
\epsilon,\, \mu \; : \; \Alt^{1} \rightarrow \Alt^{2}.
\end{eqnarray*}   
In the following we drop the sub-indices for the exterior derivatives $d_{0}$, $d_{1}$, $d_{2}.$ 
\subsection{Simple structured exterior domains}
\begin{figure}
 \psfrag{Eref}{$\MyField{E}_{\mathrm{ref}}$}
 \psfrag{Esc}{$\MyField{E}_{\mathrm{sc}}$}
 \psfrag{x}{$x$}
 \psfrag{y}{$y$}
 \psfrag{z}{$z$}
  \begin{center}
      \includegraphics[width=14cm]{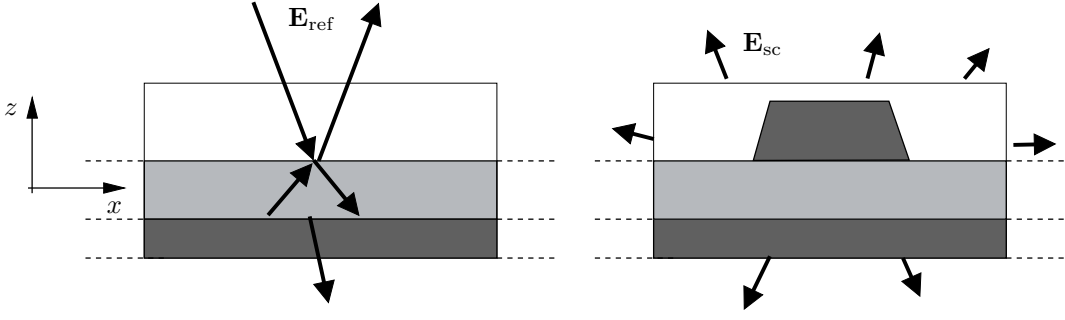}
  \end{center}
  \caption{\label{Fig:RefScatSituation} Scattering problem. For an illuminating plane wave, Maxwell's equations are solved for the reference configuration (left figure) to obtain $\VField{E}_{\mathrm{ref}}.$ The perturbed geometry with the scatterer is shown on the right hand side.  The scattered field $\VField{E}_{\mathrm{sc}}$ is the difference between the total field $\MyField{E}_{\mathrm{tot}}$ and the reference field, $\MyField{E}_{\mathrm{sc}} = \MyField{E}_{\mathrm{tot}}-\MyField{E}_{\mathrm{ref}}.$ The scattered field satisfies a radiation boundary condition. For the formulation of the scattering problem the reference field $\VField{E}_{\mathrm{ref}}$ is needed on the boundary of the computational domain. 
}
\end{figure}

We consider a situation of an isolated scatterer as in Figure~\ref{Fig:TypicalStructure}. In a scattering problem one compares a reference configuration with a perturbed configuration containing the scatterer. In Figure~\ref{Fig:RefScatSituation} the reference configuration is a wafer/mask blank with its layered media. For a prescribed illuminating plane wave the solution to Maxwell's equations in the layered media is denoted by $\MyField{E}_{\mathrm{ref}}.$ Hence the field $\MyField{E}_{\mathrm{ref}}$ already comprises the reflections at the layered media. The point here is that $\MyField{E}_{\mathrm{ref}}$ is a solution to Maxwell's equations for the reference configuration and can be directly computed. Now assume that the computational domain contains a scatterer as in Figure~\ref{Fig:RefScatSituation} (right). Actually the computational domain may contain any change of the material properties. For the prescribed illuminating plane wave we denote the total electric field by $\MyField{E}_{\mathrm{tot}}$ and introduce the scattered field as
\[
\MyField{E}_{\mathrm{sc}} = \MyField{E}_{\mathrm{tot}}-\MyField{E}_{\mathrm{ref}}.
\]
The scattered field $\MyField{E}_{\mathrm{sc}}$ is the difference between the reference field and the field observed for the perturbed geometry. This way the scattered field $\MyField{E}_{\mathrm{sc}}$ contains no incoming field components, so that a radiation boundary condition can be imposed. This yields the following formulation of the scattering problem as a coupled interior--exterior domain problem, 
\begin{eqnarray*}
d \mu^{-1}  d \VField{e} - \omega^2 \epsilon \VField{e} & = & 0 \quad \mbox{in} \; \Omega \\
d \mu^{-1}  d \VField{e}_{\mathrm{sc}}- \omega^2 \epsilon \VField{e}_{\mathrm{sc}} & = & 0 \quad \mbox{in} \; \rnum^3 \setminus \Omega \\
\VField{e} & = & \VField{e}_{\mathrm{sc}}+\VField{e}_{\mathrm{ref}} \quad \mbox{on} \; \partial \Omega \\
d \mu^{-1} \VField{e} & = & d \mu^{-1} \VField{e}_{\mathrm{sc}}+ d \mu^{-1} \VField{e}_{\mathrm{ref}} \qquad \mbox{on} \; \partial \Omega. \\
\end{eqnarray*}
Here, $\Omega$ denotes the computational domain and $\VField{e}$ the total field $\MyField{E}_{\mathrm{tot}}$ in the interior domain. The first equation is Maxwell's equation in the interior domain for the total electric field. The second equation is Maxwell's equation for the scattered electric field in the exterior domain. The last two equations are the coupling conditions on the boundary. The third equation enforces tangential continuity of the total electric field, whereas the fourth equation enforces tangential continuity of the magnetic field. As mentioned above a radiation boundary condition must be imposed for the scattered field $\VField{e}_{\mathrm{sc}}.$ We postpone the formulation of the radiation boundary condition to Section~\ref{Sec:FEMPML}.    
\subsection{Multiple structured exterior domains}
\begin{figure}
 \psfrag{Eref}{$\MyField{E}_{\mathrm{ref}}$}
 \psfrag{Esc}{$\MyField{E}_{\mathrm{sc}}$}
 \psfrag{x}{$x$}
 \psfrag{y}{$y$}
 \psfrag{z}{$z$}
  \begin{center}
      \includegraphics[width=14cm]{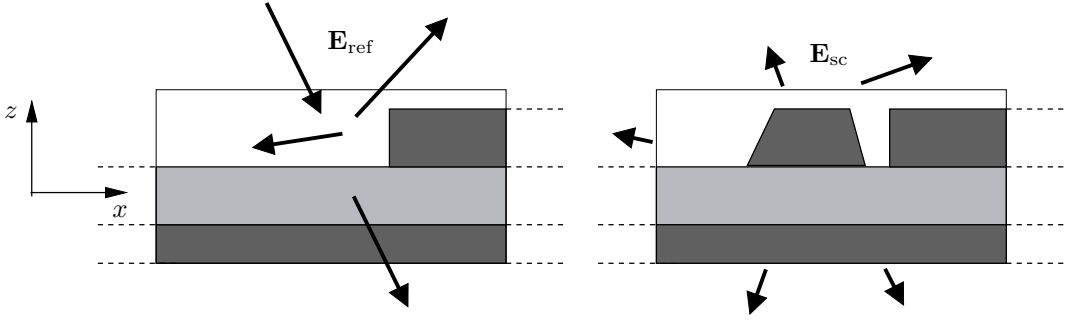}
  \end{center}
  \caption{\label{Fig:NontrivialRefScatSituation} 2D scattering problem with complicated exterior domain. The exterior domain has different layered media on the left and right side. One might use the reference geometry on the left hand side, but solving for the reference field $\MyField{E}_{\mathrm{ref}}$ is not trivial in this case.}
\end{figure}

\begin{figure}
 \psfrag{Eref1}{$\MyField{E}_{\mathrm{ref, 1}}$}
 \psfrag{Eref2}{$\MyField{E}_{\mathrm{ref, 2}}$}
 \psfrag{Esc1}{$\MyField{E}_{\mathrm{sc, 1}}$}
 \psfrag{Esc2}{$\MyField{E}_{\mathrm{sc, 2}}$}
 \psfrag{OmegaExt1}{$\Omega_{\mathrm{ext, 1}}$}
 \psfrag{OmegaExt2}{$\Omega_{\mathrm{ext, 2}}$}
 \psfrag{x}{$x$}
 \psfrag{y}{$y$}
 \psfrag{z}{$z$}
  \begin{center}
      \includegraphics[width=14cm]{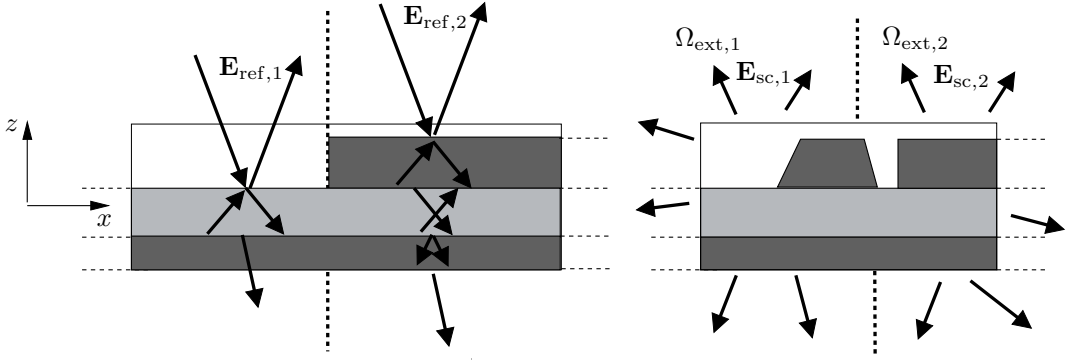}
  \end{center}
  \caption{\label{Fig:ManiRefScatSituation}Choice of different reference fields for the splitted exterior domains. $\MyField{E}_{\mathrm{ref, 1}}$ and $\MyField{E}_{\mathrm{ref, 2}}$ are solutions to Maxwell's equation in the different layered media for a prescribed illuminating plane wave. On the right hand side the exterior domain is splitted into two parts $\Omega_{\mathrm{ext, 1}}$ and $\Omega_{\mathrm{ext, 2}}.$ On each of the domain we define scattered fields  $\MyField{E}_{\mathrm{sc, 1}}$, $\MyField{E}_{\mathrm{sc, 2}}$ as the difference of the total field $\MyField{E}_{\mathrm{tot}}$ and  $\MyField{E}_{\mathrm{ref, 1}}$, $\MyField{E}_{\mathrm{ref, 2}}$ respectively. Additional coupling conditions on the boundary $\Gamma = \Omega_{\mathrm{ext, 1}} \cap \Omega_{\mathrm{ext, 2}}$ are needed.
}
\end{figure}

We now consider the more complicated situation from Figure~\ref{Fig:MoreComplicatedGeometries}. For the 2D setting we refer to Figure~\ref{Fig:NontrivialRefScatSituation}. The exterior domain in Figure~\ref{Fig:NontrivialRefScatSituation} is differently layered on the left and right horizontal half space. Hence it is not clear how to define the reference geometry. One might arbitrarily use a reference domain as given in Figure~\ref{Fig:NontrivialRefScatSituation}~(right), but this leads to a reference problem with a non--trivial solution. 
Here we propose a different approach. We horizontally split the overall space as in Figure~\ref{Fig:ManiRefScatSituation} (left) to get two reference solutions $\MyField{E}_{\mathrm{ref, 1}}$ and $\MyField{E}_{\mathrm{ref, 2}}$ in the different layered media for a prescribed illuminating plane wave. On the right hand side of Figure~\ref{Fig:ManiRefScatSituation} the exterior domain is splitted into two parts $\Omega_{\mathrm{ext, 1}}$ and $\Omega_{\mathrm{ext, 2}}.$ Now we deal with two different scattered fields defined by
\begin{subequations}
\label{DefEScat}
\begin{eqnarray}
\MyField{E}_{\mathrm{sc, 1}} & = & \MyField{E}_{\mathrm{tot}}-\MyField{E}_{\mathrm{ref, 1}} \\
\MyField{E}_{\mathrm{sc, 2}} & = & \MyField{E}_{\mathrm{tot}}-\MyField{E}_{\mathrm{ref, 2}},
\end{eqnarray} 
\end{subequations} 
where $\MyField{E}_{\mathrm{tot}}$ is again the total field. The scattered fields $\MyField{E}_{\mathrm{sc, 1}}$ and $\MyField{E}_{\mathrm{sc, 2}}$ satisfy Maxwell's equations on $\Omega_{\mathrm{ext, 1}}$ and $\Omega_{\mathrm{ext, 2}}$ respectively and are outward radiating. From the definitions~\eqref{DefEScat} we conclude that the scattered fields jump across the interface $\Gamma = \Omega_{\mathrm{ext, 1}} \cap \Omega_{\mathrm{ext, 2}}$, 
\begin{eqnarray*}
\MyField{E}_{\mathrm{sc, 2}}-\MyField{E}_{\mathrm{sc, 1}} & = & \MyField{E}_{\mathrm{ref, 1}}-\MyField{E}_{\mathrm{ref, 2}} \\
d\mu^{-1}\MyField{E}_{\mathrm{sc, 2}}-d\mu^{-1}\MyField{E}_{\mathrm{sc, 1}} & = & d\mu^{-1} \MyField{E}_{\mathrm{ref, 1}}-d\mu^{-1}\MyField{E}_{\mathrm{ref, 2}}.  
\end{eqnarray*} 
Note that this jump is exactly the difference of the reference fields. Since for both reference fields the same incoming plane wave is chosen the difference is a purely outgoing wave. This fact will play an important role, when we define transparent boundary conditions. The scattering problem with two different reference fields now reads as,  
\begin{subequations}
\label{Eqn:MultRefScat}
\begin{eqnarray}
d \mu^{-1}  d \VField{e} - \omega^2 \epsilon \VField{e} & = & 0 \quad \mbox{in} \; \Omega \\
d \mu^{-1}  d \VField{e}_{\mathrm{sc}, k}- \omega^2 \epsilon \VField{e}_{\mathrm{sc}, k} & = & 0 \quad \mbox{in} \; \Omega_{\mathrm{ext, k}}  \\
\label{Eqn:CouplingDirichletIntExt}
\VField{e} & = & \VField{e}_{\mathrm{sc}, k}+\VField{e}_{\mathrm{ref}, k} \quad \mbox{on} \; \partial \Omega \cap \partial \Omega_{\mathrm{ext, k}}  \\
\label{Eqn:CouplingNeumannIntExt}
d \mu^{-1} \VField{e} & = & d \mu^{-1} \VField{e}_{\mathrm{sc}, k}+ d \mu^{-1} \VField{e}_{\mathrm{ref}, k} \qquad \mbox{on} \; \partial \Omega \cap \partial \Omega_{\mathrm{ext, k}} \\
\label{Eqn:CouplingDirichletExtExt}
\MyField{e}_{\mathrm{sc, 2}}-\MyField{e}_{\mathrm{sc, 1}} & = & \MyField{e}_{\mathrm{ref, 1}}-\MyField{e}_{\mathrm{ref, 2}} \qquad \mbox{on} \;   \partial \Omega_{\mathrm{ext, 1}} \cap \partial \Omega_{\mathrm{ext, 2}} \\
\label{Eqn:CouplingNeumannExtExt}
d\mu^{-1}\MyField{e}_{\mathrm{sc, 2}}-d\mu^{-1}\MyField{e}_{\mathrm{sc, 1}} & = & d\mu^{-1} \MyField{e}_{\mathrm{ref, 1}}-d\mu^{-1}\MyField{e}_{\mathrm{ref, 2}} \qquad \mbox{on} \;   \partial \Omega_{\mathrm{ext, 1}} \cap \partial \Omega_{\mathrm{ext, 2}}
\end{eqnarray}
\end{subequations}
for $k=1, 2.$ The generalization to multiple reference solutions is straightforward. One only needs to account for all coupling conditions at the interfaces of two adjacent exterior domains. As a strategy one decompose the exterior domain such that the reference fields to the prescribed incoming field are numerically available. In the case of a layered media, the reference solution can be obtained quasi-analytically. For waveguide inhomogeneities as in Figure~\ref{Fig:MoreComplicatedGeometries} (left) one may also split the domain at $y=0.$ The domain with $y<0$ is a simple layered media. The domain with $y>0$ contains the waveguide (resist line) and the corresponding reference solution can be obtained by solving a 2D scattering problem on the cross-section $y=const.$ However, there is a restriction for this construction. The difference of two reference fields in the coupling condition~\eqref{DefEScat} must satisfies a radiation boundary condition defined in the next section.  

\subsection{Weak formulation}
The scattering problem~\eqref{Eqn:MultRefScat} is given in strong form. To use the finite element method one must cast the scattering problem into weak form. Since the problem is still posed on the entire space $\rnum^3$ we use test functions with local support. In the following we apply the partial integration rule for differential 1-forms $\MyField{u}, \MyField{v},$ 
\[
\int_\Omega \MyField{u} \wedge d \MyField{v}  = \int_\Omega d \MyField{u} \wedge  \MyField{v}-\int_{\partial \Omega} \MyField{u} \wedge \MyField{v}. 
\]
So given a test function $\VField{v}$ we get, 
\begin{eqnarray*}
\int_{\Omega} d \VField{v} \wedge  \mu^{-1}  d \VField{e} - \omega^2 \VField{v} \wedge \epsilon \VField{e} & = & \int_{\partial{\Omega}} \VField{v} \wedge  \mu^{-1}  d \VField{e}  \\
\int_{\Omega_{\mathrm{ext, k}}} d \VField{v}  \mu^{-1}  d \VField{e}_{\mathrm{sc}, k}- \omega^2 \epsilon \VField{e}_{\mathrm{sc}, k} & = &  \int_{\Omega_{\mathrm{ext, k}}}  \VField{v} \wedge  \mu^{-1}  d \VField{e}_{\mathrm{sc}, k}, \quad k=1, 2.   
\end{eqnarray*}
Summing  these three equations and using equations~\eqref{Eqn:CouplingNeumannIntExt},~\eqref{Eqn:CouplingNeumannExtExt} yields
\begin{eqnarray*}
\int_{\rnum^3}  d \VField{v} \wedge  \mu^{-1}  d \widetilde{\VField{e}} - \omega^2 \VField{v} \wedge \epsilon \widetilde{\VField{e}} & = &
\int_{\partial{\Omega} \cap \partial \Omega_{\mathrm{ext, 1}}}  \VField{v} \wedge  \mu^{-1}  d \VField{e}_{\mathrm{ref}, 1}+
\int_{\partial{\Omega} \cap \partial \Omega_{\mathrm{ext, 2}}}  \VField{v} \wedge  \mu^{-1}  d \VField{e}_{\mathrm{ref}, 2}+ \\
{} & {} & \int_{\partial{\Omega_{\mathrm{ext, 1}}} \cap \partial \Omega_{\mathrm{ext, 2}}}  \VField{v} \wedge  \mu^{-1}  d \left(\VField{e}_{\mathrm{ref}, 2}-\VField{e}_{\mathrm{ref}, 1} \right),
\end{eqnarray*}
where $\widetilde{\VField{e}}$ denotes the compound field $\widetilde{\VField{e}} = \VField{e}+\VField{e}_{\mathrm{sc}, 1}+\VField{e}_{\mathrm{sc}, 2}.$ This equation accounts for the coupling of the Neumann data on the various boundaries. But $\widetilde{\VField{e}}$ jumps across the boundaries, so that it is not feasible in a finite element discretization. In a further step we therefore must incorporate the Dirichlet coupling conditions~\eqref{Eqn:CouplingNeumannIntExt} and~\eqref{Eqn:CouplingNeumannExtExt}. To do that one adds supplemental fields with local support near the interface boundaries to $\widetilde{\VField{e}}$ making the so constructed field $\VField{e}'$ continuous across the boundaries. The choice of the additional fields is not unique, but must account for the construction of transparent boundary conditions in the next section. The corner point $\partial \Omega \cap \partial{\Omega_{\mathrm{ext, 1}} \cap \Omega_{\mathrm{ext, 2}}}$ needs a special treatment for a proper matching of the Dirichlet values at this point. Let us denote by $B_\epsilon$ a small vicinity of this corner point. We define a smooth function $\chi$ with the following properties, 
\begin{eqnarray*}
\chi & = & 1 \quad \mbox{on}\; \partial \Omega \\
\chi & = & 0 \quad \mbox{on}\; \partial{\Omega_{\mathrm{ext, 1}} \cap \partial \Omega_{\mathrm{ext, 2}}} \setminus B_\epsilon.
\end{eqnarray*}
Now we introduce fields $\VField{e}_{\mathrm{ref}, 1}'$ and $\VField{e}_{\mathrm{ref}, 2}'$ such that
\begin{eqnarray*}
\VField{e}_{\mathrm{ref}, 1}' & = & \chi \VField{e}_{\mathrm{ref}, 1} \\
\VField{e}_{\mathrm{ref}, 2}' & = & \chi \VField{e}_{\mathrm{ref}, 2}+(1-\chi) \left( \VField{e}_{\mathrm{ref}, 2}-\VField{e}_{\mathrm{ref}, 1} \right).
\end{eqnarray*} 
This allows for the definition
\[
\VField{e}' = \widetilde{\VField{e}}+\VField{e}_{\mathrm{ref}, 1}'+\VField{e}_{\mathrm{ref}, 2}',
\]
which is continuous across the coupling boundaries. The variational problem for $\VField{e}'$ now reads as
\begin{eqnarray}
\label{Eqn:WeakMultRefScat}
\nonumber
\int_{\rnum^3}  d \VField{v} \wedge  \mu^{-1}  d \VField{e}' - \omega^2 \VField{v} \wedge \epsilon \VField{e}' & = &
\int_{\partial{\Omega} \cap \partial \Omega_{\mathrm{ext, 1}}}  \VField{v} \wedge  \mu^{-1}  d \VField{e}_{\mathrm{ref}, 1}+
\int_{\partial{\Omega} \cap \partial \Omega_{\mathrm{ext, 2}}}  \VField{v} \wedge  \mu^{-1}  d \VField{e}_{\mathrm{ref}, 2}+ \\
\nonumber
{} & {} & \int_{\partial{\Omega_{\mathrm{ext, 1}}} \cap \partial \Omega_{\mathrm{ext, 2}}}  \VField{v} \wedge  \mu^{-1}  d \left(\VField{e}_{\mathrm{ref}, 2}-\VField{e}_{\mathrm{ref}, 1} \right)+ \\
\nonumber
{} & {} & \int_{\Omega_{\mathrm{ext, 1}}}  d \VField{v} \wedge  \mu^{-1}  d \VField{e}_{\mathrm{ref}, 1}' - \omega^2 \VField{v} \wedge \epsilon \VField{e}_{\mathrm{ref}, 1}'+ \\ 
{} & {} & \int_{\Omega_{\mathrm{ext, 2}}}  d \VField{v} \wedge  \mu^{-1}  d \VField{e}_{\mathrm{ref}, 2}' - \omega^2 \VField{v} \wedge \epsilon \VField{e}_{\mathrm{ref}, 2}'. 
\end{eqnarray}
The construction of $\VField{e}'$ looks very tricky in this analytic setting, but is very simple in the finite element context. An explicite construction of the function $\chi$ can be avoided. Instead one imposes the Dirichlet jump conditions directly on the finite element coefficient vector.
\section{Transparent Boundary Conditions/Perfectly Matched Layers}   
\label{Sec:FEMPML}
So far, in all considerations the various scattering problems were posed on
the entire domain $\rnum^3$ and are therefore numerically not feasible. This is
overcome by using transparent boundary conditions. We use the perfectly
matched layer method introduced by Berenger \cite{Berenger:94, Chew:94}. This method
exploits the analytic continuation properties of the scattered field in the
exterior domain. In a nutshell using an appropriate complex continuation, the
scattered field is transformed to an exponentially decaying field without
affecting the matching condition with the field in the interior domain. This allows for a truncation of the computational domain. In an
earlier paper we proposed an extremely efficient adaptive PML method which also copes with various kinds of inhomogeneous exterior domains \cite{Zschiedrich2005a, Zschiedrich2006a}. The method can also be applied to~\eqref{Eqn:WeakMultRefScat}. One derives a weak form of Maxwell's equation with PML which has exactly the same form as equation~\eqref{Eqn:WeakMultRefScat}, but where the fields $\VField{e}_{\mathrm{ref}, 1}$, $\VField{e}_{\mathrm{ref}, 2}$,  $\VField{e}_{\mathrm{sc}, 1}$ and  $\VField{e}_{\mathrm{sc}, 1}$ are replaced by their complex continuation. The last four terms on the right hand side of equation~\eqref{Eqn:WeakMultRefScat} are non-standard. These terms model the coupling of the illuminating field with non-trivial structures in the exterior domain. Within the PML the magnitude of these further right hand side terms is proportional to $\|\VField{e}_{\mathrm{ref}, 2}-\VField{e}_{\mathrm{ref}, 1}\|.$ As far as $\VField{e}_{\mathrm{ref}, 1}$ and $\VField{e}_{\mathrm{ref}, 2}$ share the same incoming field, we conclude that the additional terms are also evanescent within the PML. This must be guaranteed when splitting the exterior domain into different reference geometries.

\begin{figure}
 \psfrag{cd_l}{$l_{\mathrm{cd}} = 310 \mathrm{nm}$}
 \psfrag{h}{$h_{\mathrm{mosi}} = 65 \mathrm{nm}$}
 \psfrag{x}{$x$}
 \psfrag{y}{$y$}
 \psfrag{z}{$z$}
  \begin{center}
      \includegraphics[width=7cm]{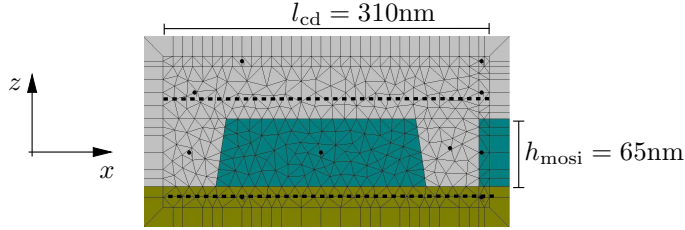}
  \end{center}
  \caption{\label{Fig:TestProblem} Test problem with different right and left hand side exterior domains. A MoSi structure ($n_{\mathrm{mosi}} = 2.52 + 0.596i$) is mounted on a glass substrate ($n_{\mathrm{g}} =  1.5306$). The initial finite element mesh is shown. Quadrilaterals are modeled as infinitely prolonged by using the adaptive PML method. The structure is illuminated by a $s$-polarized plane wave ($E_x, E_z = 0$)  with vaccum wavelength $\lambda = 193 \mathrm{nm}$ and wave direction $\hat{k} = (-1/\sqrt{2}, 0, -1/\sqrt{2}).$ To compare the simulated field with the periodic setting in Figure~\ref{Fig:TestProblemPeriodic} the field values are extracted on the dashed lines.  
}
\end{figure}

\begin{figure}
 \psfrag{cd_l}{$l_{\mathrm{\pi}}$}
  \begin{center}
      \includegraphics[width=10cm]{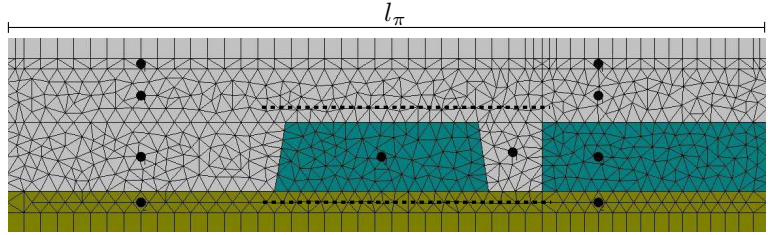}
  \end{center}
  \caption{\label{Fig:TestProblemPeriodic} Same setting as in Figure~\ref{Fig:TestProblem} but periodic boundary conditions are used in horizontal direction. The periodicity length $l_{\pi}$ was gradually increased to suppress the cut-off error caused by the artificial periodification. Field values are extracted on the dashed lines placed as in Figure~\ref{Fig:TestProblem}.
}
\end{figure}

\section{Imaging}
In this section we briefly sketch how to extract the far field information. We first refer to Figure~\ref{Fig:RefScatSituation} with of a simple layered media in the exterior domain. A rigorous field evaluation formula of the scattered field $\VField{e}_{\mathrm{sc}}$ in the upper half space can be derived by means of the Rayleigh-Sommerfeld integral~\cite{Singer:05a}. In microlithography application one is mainly interested in the far field data. Moving an evaluation point towards infinity in a direction $\tau (nx, ny, nz), \tau \rightarrow \infty,$ one shows that the field value converges up to a scaling to the Fourier transform of the scattered field in the $xy-$ plane.   

Alternatively to an evaluation of the Rayleigh-Sommerfeld integral, one may use Green's tensor $\MyField{G}_{\vec{x}_p},$
\[ d \mu^{-1} d \MyField{g}_{\vec{x}_p} -\omega^2 \epsilon \MyField{g}_{\vec{x}_p} = \delta(\cdot-{\vec{x}_p}), \quad \mbox{in}\; \rnum^3 \setminus \Omega 
\]
in the exterior domain to evaluated the field in the exterior domain. One gets
\[
\VField{e}_{\mathrm{sc}}(\vec{x}_p) = \int_{\partial{\Omega}} \VField{e}_{\mathrm{sc}} \mu^{-1} d \MyField{g}_{\vec{x}_p} - \MyField{g}_{\vec{x}_p} \mu^{-1} d \VField{e}_{\mathrm{sc}},
\]
so that the scattered field is only needed on the boundary of the computational domain. Since typically one is only interested in the far field, it is sufficient to use the asymptotics of the Green function for $\vec{x}_p = \tau (nx, ny, nz),$ $\tau \rightarrow \infty,$ which can be easily computed for the layered media. The situation for the more complicated exterior domain in Figure~\ref{Fig:NontrivialRefScatSituation} is more involved. The computed scattered field $\VField{e}_{\mathrm{sc}}$ depends on the choice of reference field $\VField{e}_{\mathrm{ref}}.$ The reference field already comprises reflected fields from the infinite resist line.  From a metrology point of view this corresponds to a calibration. Once a reference field in the entire exterior domain $\rnum^3\setminus \Omega$ as in Figure~\ref{Fig:NontrivialRefScatSituation} (left) is fixed, the scattered field $\VField{e}_{\mathrm{sc}} = \VField{e}_{\mathrm{tot}}-\VField{e}_{\mathrm{ref}}$ has a purely continuous spectrum and can be computed by means of either a Rayleigh-Sommerfeld integral or by using of Green's tensor asymptotics. The computation of the Green function's asymptotics requires a further numerical simulation of a scattering problem for each far field evaluation point. We detail this in a later paper. 

\begin{figure}
  \begin{center}
      \includegraphics[width=10cm]{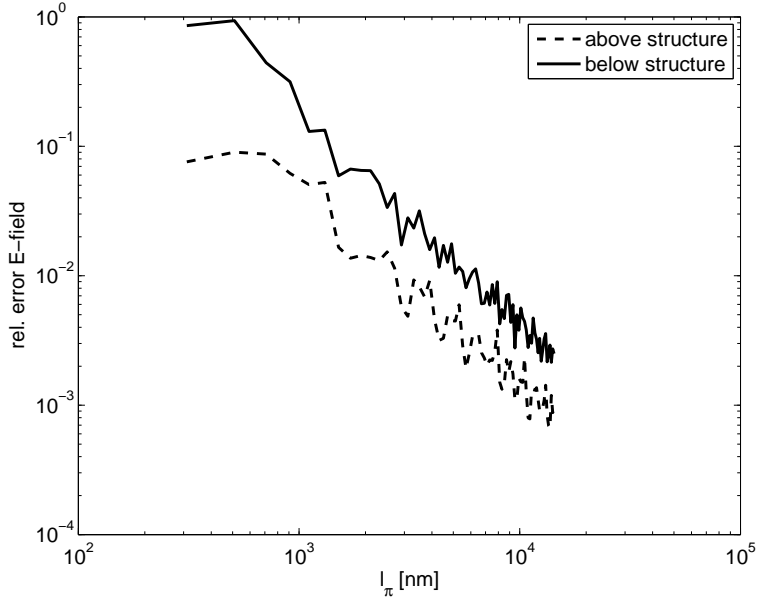}
  \end{center}
  \caption{\label{Fig:TestProblemConvergence} Convergence of the periodic solution with increasing periodicity length $l_\pi$ to the solution obtained with the new method. The field values were extracted on horizontal lines above and below the structure as in Figures~\ref{Fig:TestProblem},~\ref{Fig:TestProblemPeriodic}. Even for moderate accuracy demands one observes a tremendous speed-up by the new method compared to the periodic setting.
}
\end{figure}

\section{Numerical Example}
We applied the new method to the simple 2D test problem given in Figure~\ref{Fig:TestProblem}. An analytic solution to this problem is not known. To verify the accuracy of the method we simulated the structure on an artificially periodified domain as in Figure~\ref{Fig:TestProblemPeriodic}. When increasing the periodicity length $l_{\pi}$ one expects convergence to the solution obtained for the isolated case. To measure the error we extracted the electric field values on horizontal lines below  and above the structure (Figures~\ref{Fig:TestProblem},~\ref{Fig:TestProblemPeriodic}). As can be seen from Figure~\ref{Fig:TestProblemConvergence} the field values for the periodic setting converges to the values obtained with the new method confirming our theory of scattering within multiple exterior domains. Even for moderate accuracy demands one observes a tremendous speed-up by the new method compared to the periodic setting.     
\bibliography{lit}     
\bibliographystyle{spiebib}   
\end{document}